# Magnetization Vorticity and Exchange Bias Phenomena in Arrays of Small Asymmetric Magnetic Rings

Deepak K Singh,[a] T. Yang and Mark T Tuominen

*Department of Physics, University of Massachusetts Amherst*

Arrays of nanoscopic magnetic asymmetric rings, 150 nm in outer diameter, are fabricated using the techniques of electron-beam lithography, angular deposition and ion-beam etching. Magnetic measurements for cobalt asymmetric rings at room temperature verifies previous reports of vortex magnetic state formation of a desired circulation direction for the application of external magnetic field along the asymmetry axis of the rings. However, the main theme of this article is the observation of exchange bias phenomena when the ring samples are cooled down to low temperature in the presence of a positive magnetic field. Very interestingly, the observed exchange bias effect is negative for along and perpendicular orientations of ring's asymmetry axis with respect to the in-plane external magnetic field. This is in good quantitative agreement with the random interface model proposed by Malozemoff et al. For the application of inplane external magnetic field at 45 degree with respect to the asymmetry axis, the exchange bias effect is positive. Unlike the exchange bias effects in thin films, this is a very unusual observation indicating that exchange bias phenomena of opposite natures can be manipulated by appropriate combinations of geometrical constraint and external magnetic field direction, in addition to the interfacial interactions between ferromagnetic (FM) and antiferromagnetic (AFM) layer.

---

[a] Present address: dsingh@nist.gov

# 1. INTRODUCTION

In recent times nanoscopic magnetic rings are emerging as important candidate for magnetic memory device applications mostly because of the formation of a vortex magnetization state,[1] near zero field value, and its associated zero net magnetization. For potential applications of rings as memory elements, only creating a vortex magnetic state is not sufficient but controlling the direction of vortex circulation is also necessary. Also it has been found that in the case of small (outer diameter of the order of 100 nm) symmetric magnetic ring structures, the magnetic transition process is statistically dominated by the rotation of domain walls instead of vortex magnetic state formation.[2] On the other hand if the rings are asymmetric, the probability of vortex formation can be greatly enhanced by creating a minimum energy location in the ring structure. Asymmetricity in the ring structure not only increases the possibility of vortex formation but also helps control the vortex circulation direction. There are several ways to control the vortex circulation direction e.g. by using well-timed pulses of radial field application[3] and a simple magnetic field application.[4] Perhaps the most relevant is the uniform magnetic field application, as it is easy to implement. Klaui et al. have performed micromagnetic simulations for an asymmetric ring where ring's width changes slowly from one side to the other, confirming that by creating asymmetry, one can obtain a vortex state of desired circulation direction using the application of simple magnetic field.[1]

In this article, we report magnetic measurements for arrays of small (150 nm outer diameter) asymmetric cobalt ring samples. Measurements at room temperature (300 K) for asymmetric ring arrays verify previous experimental observations of vortex magnetic state formation for in-plane application of magnetic field. In addition, strong evidence of vortex magnetization state formation near zero field value is also observed in the magnetic measurements for perpendicular orientation of magnetic field, with respect to the ring plane. When the asymmetric ring sample is cooled down to low temperature (2 K) in the presence of an in-plane cooling field, exchange bias phenomena are observed. Furthermore, as the orientation of asymmetry axis with respect to (in-plane) external magnetic field is changed, exchange bias effects of both negative and positive shifts ($H_E$, exchange field) are observed. In general, an exchange bias phenomenon arises as a result of interfacial interaction between a ferromagnetic layer and the neighboring anti-ferromagnetic layer.[5] This interfacial interaction creates a unidirectional anisotropy at the interface,[6, 7, 8] which causes a shift of the magnetic hysteresis loop towards negative (positive) direction of field, called negative (positive) exchange bias. Earlier Ambrose et al. had performed a detail measurement to demonstrate the angular dependence of exchange bias coupling for a FM/AFM bilayer film.[9] In another work,[10] strong exchange bias effects were found to be varying in Co/CoO film as a function of the Co layer thickness. Significantly enough, the AFM film layer (CoO) in this case is thick enough to remain mostly uncompensated. In our case, preliminary quantitative investigations show that it is the polycrystalline nature of magnetic material, Co in this case, that creates a random magnetic field at the interface of grainy Co (FM) layer and the native oxide, CoO, anti-ferromagnetic (AFM) layer that leads to negative exchange bias effect. Usually positive exchange bias is observed if the cooling field value is very high[11] (>> 1 *Tesla*) but in our measurements we have used the same cooling field value. Observation of exchange bias in the case of magnetic disk has been also reported recently.[12]

## 2. FABRICATION AND MAGNETIZATION MEASUREMENTS

Asymmetric rings are fabricated using our new fabrication method that involves the techniques of electron-beam lithography, angular deposition and ion-beam etching. A detailed description of this new fabrication technique can be found elsewhere.[13] After using electron-beam lithography to fabricate an array of empty pores in PMMA, of thickness 70 nm ($h$) and diameter 150 nm ($w$) with lattice constant of 400 nm, the technique of masked angular deposition is used to make small asymmetric nanometer scale rings. The critical angle for masked angular deposition in this case is given by $\theta = \tan^{-1}\left(\frac{w}{h}\right) = \theta_{crt}$ ~ 64 degree. To make the width distribution of deposited material asymmetric across the nanoscopic pores wall, we control the rotation of substrate holder while material deposition continues. In this regard, the substrate is uniformly rotated for one third of evaporation time and then rotation is stopped for another one third of evaporation time while the substrate is still mounted at the same angle (so that only walls of the pores are exposed) and at last we rotate the substrate again for the remaining one third of evaporation time. As a result, one side of the ring arm becomes wider as compared to the other side. The desired ring arm widths in our experiment are 30 nm on the wider side and 20 nm on the thinner side of the ring arm. After depositing cobalt material onto the walls of the pores, ion-beam etching is used to remove the undesired material from on top of the PMMA film as well as from the bottom of the pores. A calibrated ion-beam etching rate for cobalt material is used to tune the thickness of the rings which is 20 nm in this case. After ion-beam etching, the sample is cleaned in acetone solvent to remove the remaining PMMA. Thus arrays of polycrystalline cobalt asymmetric rings are obtained. Samples are characterized using scanning electron microscope (Figure 1).

Magnetic measurements were performed in a SQUID magnetometer at both room temperature (300 K) and low temperature (2 K) for various in-plane and perpendicular orientations of field with respect to the asymmetry axis (as shown in Fig. 1 with thick straight line arrows) of the sample. Magnetic measurements for asymmetric rings are shown in Figures 3, 4 and 5. In asymmetric rings, the ring width (or thickness or both) vary along the circumference, with their minima and maxima separated by $180°$. If the initial magnetic field is along the symmetry axis, one domain wall will be generated at the thinnest location and the other domain wall at the thickest location after the field is reduced from saturation value. Recently Zhu et al.[2] have shown that in the presence of a reversal magnetic field, these two domain walls still have two equivalent directions to move, and the situation is not very different from that of the symmetric rings. In that case, the probability for vortex formation is not very different from that of a symmetric ring and is less probable than the rotation of domain walls. On the other hand, by using the application of a field directed along the asymmetry axis, the direction of movements of domain walls can be controlled towards minimum energy configuration which leads to the formation of vortex magnetic state of desired circulation direction (Figure 2).[1,2] Only by the application of a magnetic field, in a direction different from the symmetry axis (positive or negative angle), a specific CCW (counter-clockwise) or CW (clockwise) vortex magnetic state can be obtained.

In Figure 3, in-plane magnetic measurements data for asymmetric rings at 300 K are shown. For magnetic field application along the asymmetry axis, we clearly see an abrupt change in the magnetization around -120 Oe of field value. This is marked by a red arrow in the figure. Similar behavior has been observed previously in the case of asymmetric rings arrays[2]. The abrupt change in magnetization is identified with the vortex magnetic state formation. In this case the circulation direction of vortex is CCW, as explained before. When the magnetic field is applied at 45 degree from the asymmetry axis, again we see the signature of vortex state formation around -120 Oe of field value. As discussed earlier, the probability of vortex formation in this case is smaller compared to the field application along the asymmetry axis. When the in-plane field is applied perpendicular to the asymmetry axis of the ring then we do not see any plateau near the expected vortex transition. It indicates no vortex formation in this case. It can be argued that in this case asymmetric ring is no different than symmetric ring and magnetic transition may have occurred via the rotation of domain walls only. These observations are consistent with the report of Zhu et al.[2]

We have also performed the room temperature (300 K) magnetic measurement of asymmetric rings arrays for perpendicular application of magnetic field with respect to ring's plane and are plotted in Figure 4. As we can see in this figure, a small plateau of magnetization develops around 500 Oe of magnetic field, similar to in-plane magnetic field measurements. For perpendicular application of field, initially magnetic spins are saturated along the field application direction for very high value of field. As the field is swept to very high negative value, magnetization saturation direction is reversed (as shown in the Figure 4 by green arrows). In the process of reversing the saturation directions, the magnetic spins go through the plane of the asymmetric ring. When the magnetic spins are in-plane then because of minimum energy configuration of vortex magnetic state near zero field value, it is possible that asymmetric ring acquires the vortex magnetic state configuration. Interestingly, the switching field value (~ 500 Oe) for the onset of vortex magnetic state is quite large in this case. More experimental and theoretical works are necessary to further understand this behavior. Unlike the in-plane magnetic field case, the vortex circulation direction cannot be controlled in purely perpendicular field application case.

Low temperature (2 K) magnetic measurements of asymmetric ring for in-plane field applications are shown in Figure 5. In this figure, the red curve represents the magnetic measurement for external field application along the symmetry axis, blue curve for the field application along the asymmetry axis and green curve for the field application at 45 degree from the asymmetry axis. Careful observations of these data reveal that the whole hysteresis curves are shifted along the field axis in each case. Shifting of the hysteresis curves along the field axis indicate the exchange bias effect at low temperature.[14, 15] In the measurement process, asymmetric ring sample was cooled from room temperature in the presence of an external in-plane magnetic field (cooling field, $H_{cf}$) of 5000 Oe. Following this process, the hysteresis curve measurements were started at 2 Tesla field value. This is a typical measurement procedure for exchange bias measurements. As we see in Figure 5, red and blue curves are shifted along the negative values of the field while green curve is shifted along the positive value of the field. The shift in the magnetic hysteresis loop, also called the exchange field $H_E$, is quantitatively determined using $|H_E| = \left( \dfrac{H_{right} + H_{left}}{2} \right)$, where $H_{right}$ and $H_{left}$ are the field values at which $M = 0$. The values of $H_E$ for red and blue curves are almost similar and are about -700 Oe. For green curve, this value is about 400 Oe along the positive direction of magnetic field.

## 3. DISCUSSION

As mentioned before, exchange bias usually arises from the interaction between a ferromagnetic (FM) material, like Co, and an antiferromagnetic (AFM) material, like CoO. In this case, CoO is the native oxide layer of Co material. Native oxide layer forms due to the natural oxidation process and is usually 3-4 nm thick.[16] In general the exchange bias shift is negative i.e. $H_E < 0$ for a positive cooling field. However, it has been found that samples exposed to large positive cooling fields ($H_{cf} >> 1$ Tesla) can exhibit positive exchange bias also i.e. $H_E > 0$.[17] For polycrystalline systems (our system), a typical interface between FM and AFM layers is discussed by Berkowitz et al.[15] They shown that such interface involves random distribution of magnetic particles and thus lead to frustrated exchange bonds. A theoretical model to explain the exchange bias effects in polycrystalline systems was proposed by Malozemoff.[18] According to this model, also called random interface model, random interface roughness gives rise to a random magnetic field that acts on the interface spins, yielding unidirectional anisotropy. The later causes the asymmetric offset of the hysteresis loop in negative direction of magnetic field. According to Malozemoff, the shift $H_E$ of the hysteresis loop is given by

$$H_E = \frac{2}{M_F t_F} \sqrt{\frac{J_A K_A}{a}},$$

where $M_F$ and $t_F$ are the respective saturation magnetization and thickness of ferromagnetic layer, $a$ is the lattice constant and $J_A$ and $K_A$ are the respective exchange and anisotropy constants. The domain wall can be either in ferromagnetic or antiferromagnetic side and $K_A$ is lesser of two anisotropies. Above expression can be further simplified as

$$H_E = \frac{2K_A}{M_F t_F}\sqrt{\frac{J_A}{aK_A}} \approx \frac{1}{t_F} \times \frac{2K_A}{M_F} \times d_w.$$

Here $d_w$ is the width of domain wall in FM or AFM layer, whichever is smaller, and is of the order of 18 nm.[19] Now, we calculate the value of $H_E$ for our experiment using $H_A = \frac{2K_A}{M_F} \approx 1000$ Oe for polycrystalline cobalt material[20] and $t_F = 30$ nm (the arm width on thicker side of ring), we get

$$H_E = \frac{1000 \times 18}{30} = 600 \text{ Oe}.$$

This value is very close to the experimental data of blue and red curves of asymmetric rings measurements at 2 K (Figure 5). However it does not explain the positive exchange bias effect exhibited by green curve in the experimental data.

Recently Kiwi et al.[7] proposed a model, called "the frozen interface model", to explain the positive exchange bias (PEB) phenomena observed in FM/AFM interfaces. Basic assumption in this model is the spin glass like rigidity of the antiferromagnetic monolayer near the interface, as the sample is cooled down below $T_N$ in a very high positive magnetic field. In that case, ferromagnet is completely saturated and antiferromagnet spins are fixed, except those at the interface monolayers. Therefore the only energy difference arises due to the change of interface configurations. Minimization of total energy gives condition for the crossover from negative exchange bias to positive exchange bias as a function of cooling field $H_{cf}$:

$$H_{cf} > h_{EB} \text{ where } h_{EB} = \frac{2|J_{F/AF}|}{\mu_B g_{AFM}}.$$

In the above expression, $J_{F/AF}$ is the interfacial exchange constant, $\mu_B$ is Bohr Magneton and $g_{AFM}$ is gyromagnetic ratio for antiferromagnet. The experimental data of green curve in Fig. 5 was obtained using the $H_{cf}$ value of 0.5 Tesla. For positive exchange bias effect to occur, $H_{cf} > h_{EB}$ and therefore the upper limit for $h_{EB}$ in this case is 0.5 T. We use this upper limit value of $h_{EB}$ to calculate the interfacial exchange constant $|J_{F/AF}|$ using above expression. For this purpose, we take $g_{AFM} \sim 2$.[21] The estimated value of $|J_{F/AF}|$ comes out to be ~ 0.3 meV. In a similar magnetic system with interfacial layers Fe (FM)/FeF$_2$ (AFM) which exhibits positive exchange bias effect, the calculated value of $|J_{F/AF}|$ has been found to be in the range of ~ 1 meV.[22] Thus the interfacial exchange constant in the present case is smaller compared to a similar magnetic system Fe/FeF$_2$. Estimated interfacial exchange constant (0.3 meV) is possibly limited by the size of polycrystalline cobalt grains. As mentioned earlier in the fabrication section, the rings are made of polycrystalline Co material and the average Co grain size is ~ 1 nm. In general, the exchange length for crystalline cobalt material is of the order of ~ 3.8 nm.[23] Therefore, smaller size Co grains would limit the exchange length and hence the interfacial exchange constant. Since $|J_{F/AF}|$ is directly proportional to the cooling field so PEB is observed at small cooling field of 0.5 T.

In brief, we observe exchange bias phenomena of both negative and positive natures in the magnetic measurements for asymmetric rings arrays. Based on the preliminary quantitative analysis, we have found that the negative exchange bias effect is arising due to the random interfacial interaction of FM/AFM layers and the positive exchange bias is resulting from the spin glass like rigidity of the AFM monolayer near the interface. Qualitatively, these behaviors arise due to the asymmetric nature of magnetic rings. Further experimental and theoretical investigations are strongly desirable to understand the role of asymmetricity in tuning the nature of exchange bias phenomena.

ACKNOWLEDGMENTS

This project was supported by NSF Grants DMR-0531171, DMR-0306951 and MRSEC.

**Figure Captions**

**Figure 1.** (a) SEM image of asymmetric ring arrays. For imaging purposes, rings were fabricated out of silver metal. Outer diameter of the asymmetric rings are about 150 nm and the center-to-center distance is about 400 nm. Widths of the rings are 20 nm at the thinnest location and 30 nm at thickest location. (b) SEM image of single asymmetric ring. In this picture, we can clearly see the asymmetricity in the width of the ring. White straight line arrow indicates "asymmetry" axis while the dashed arrow indicates "symmetry" axis of the asymmetric ring.

**Figure 2.** A schematic illustration, based on Klaui et al. (reference 1) micromagnetic simulation for asymmetric ring, of desired vortex state formation (in counterclockwise direction) only by the application of simple magnetic field at positive angle. Asymmetry in the ring's width creates a minimum potential state at the thinnest location and DWs have a tendency to slide down to this minimum energy location.

**Figure 3.** Magnetic measurements data for asymmetric rings at 300 K. In this figure, magnetic measurements are shown for three different in-plane directions of magnetic fields. Red arrows indicate the onset of vortex formation while a red arrow with black cross, last figure, indicates that vortex formation is not likely.

**Figure 4.** Room temperature magnetic measurement data of asymmetric ring for perpendicular orientation of magnetic field with respect to the ring's plane. In this case also the formation of vortex magnetic state is possible (as indicated by red arrow).

**Figure 5.** This figure shows the low temperature (2 K) magnetic measurements data of asymmetric ring arrays for in-plane application of magnetic field. The ring sample is field cooled ($H_{cf}$ ~ 5000 Oe) before running the magnetic hysteresis measurement. The red curve represents the magnetic measurement for external field application along the symmetry axis, blue curve for the field application along the asymmetry axis and green curve for the field application at 45 degree from the asymmetry axis.

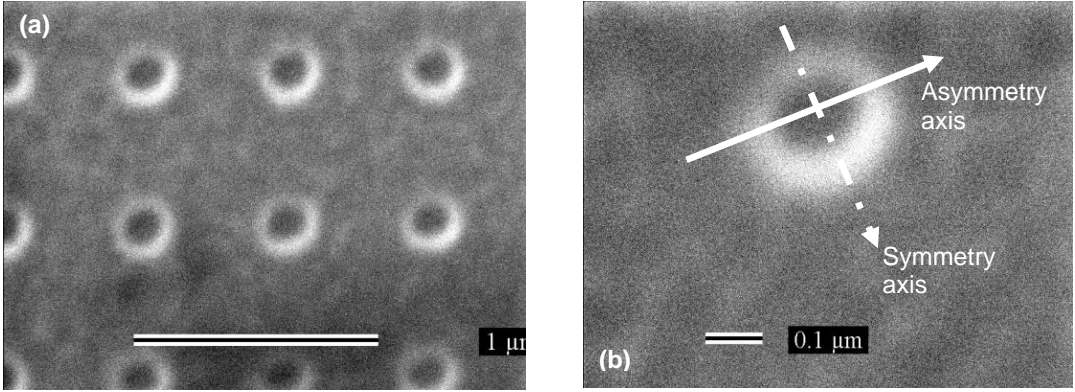

**Figure 1**

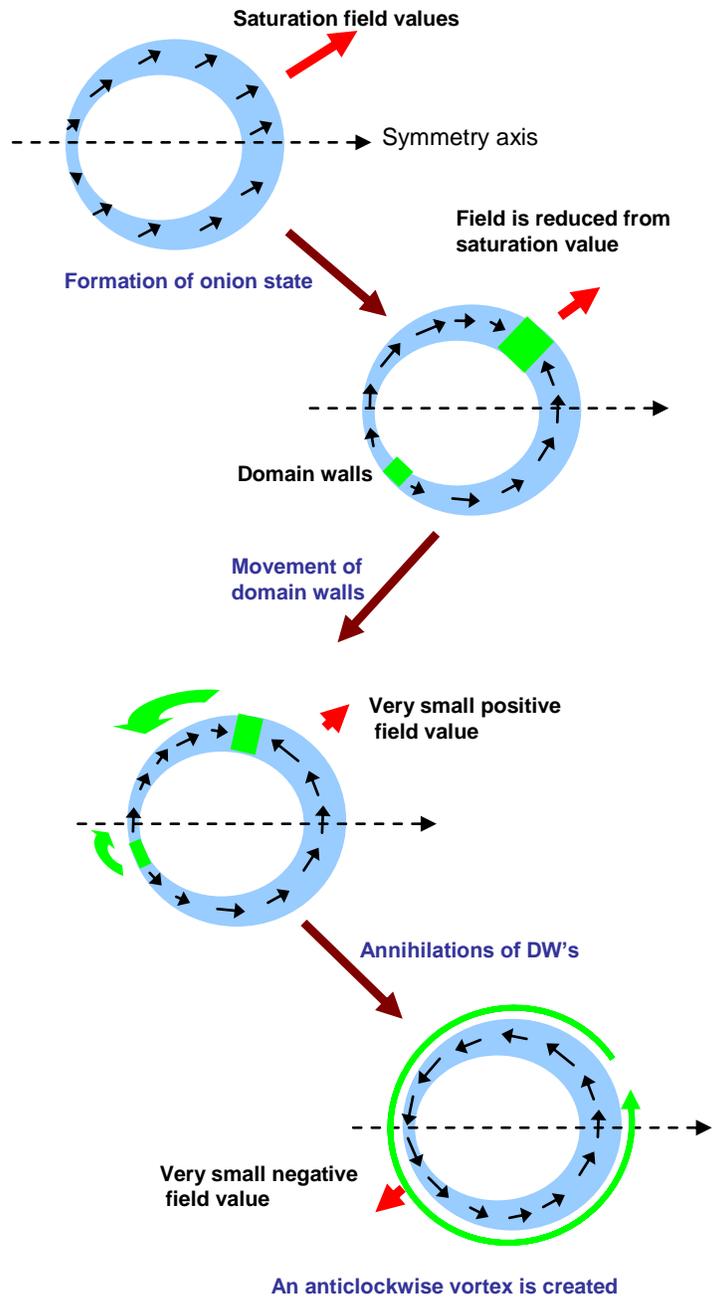

**Figure 2**

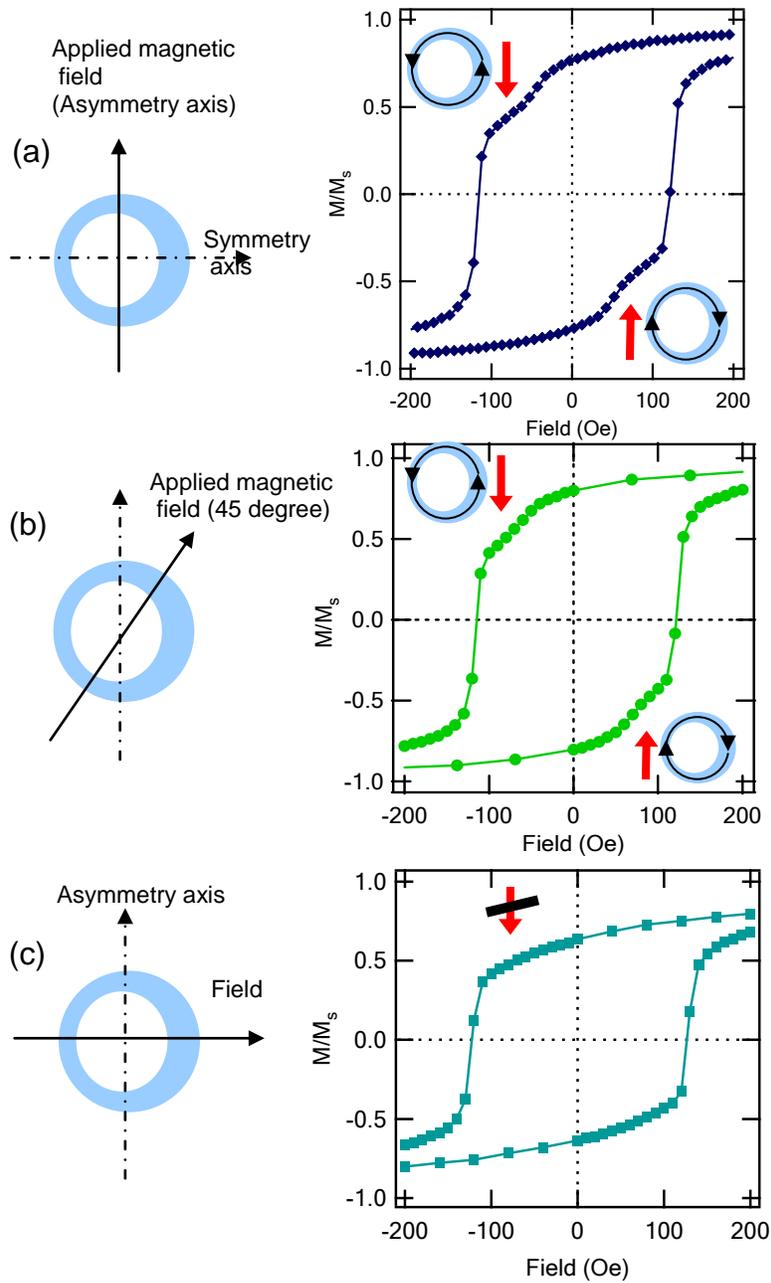

**Figure 3**

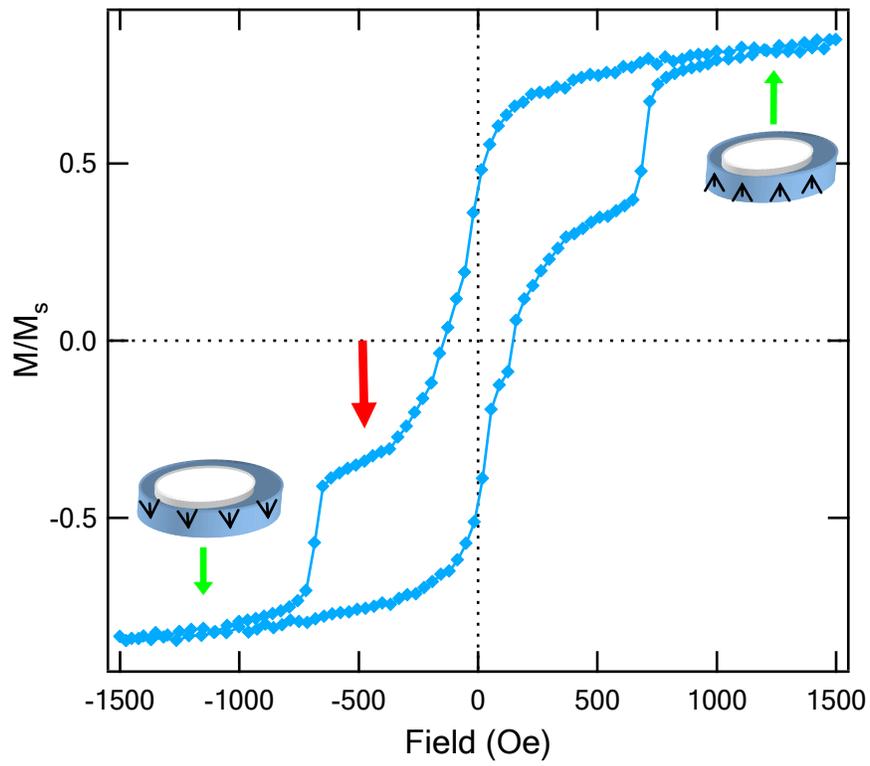

**Figure 4**

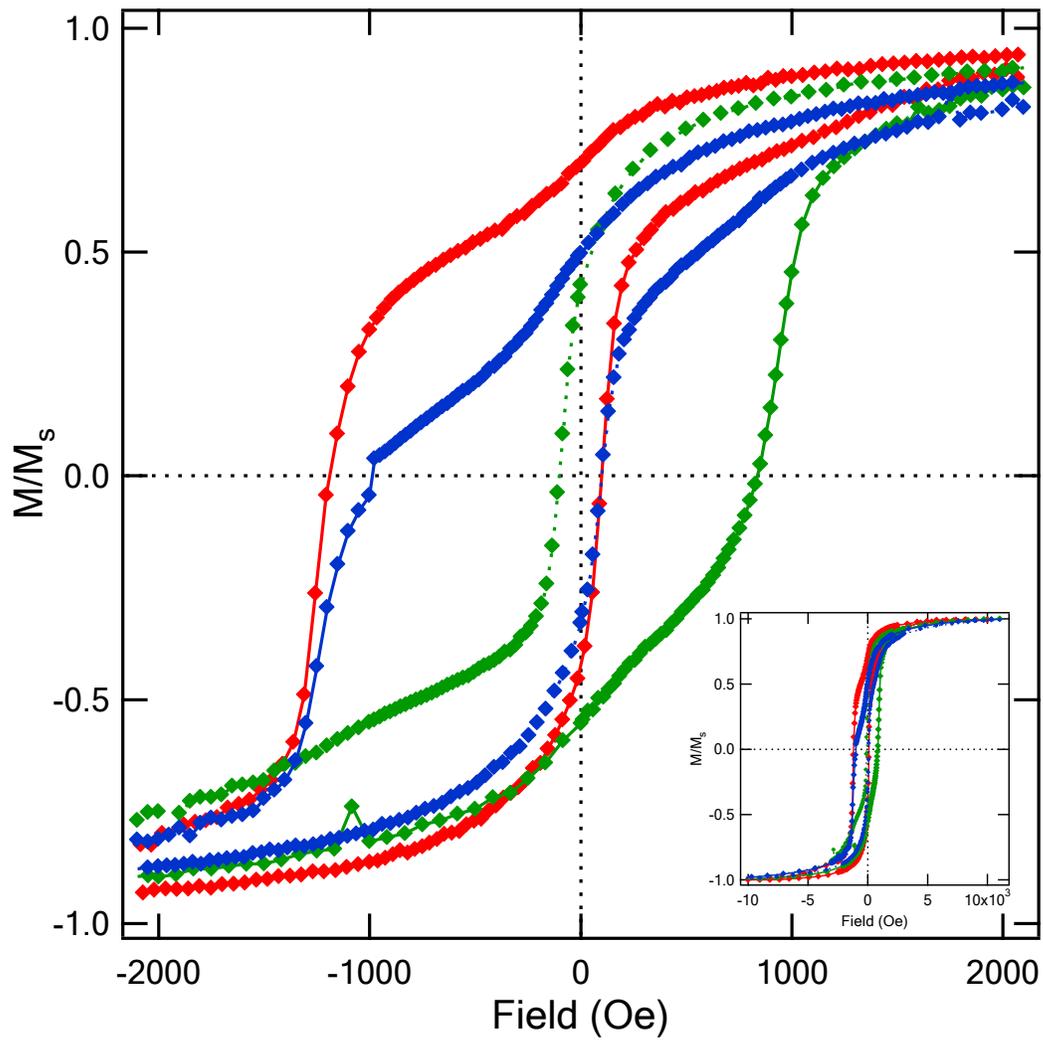

**Figure 5**

**References**

[1] M. Klaui, C A F Vaz, L. Lopez-Diaz and J A C Bland, *J. Phys.: Condens. Matt.* **15,** R985 (2003)

[2] F. Zhu, G. Chern, O. Tchernyshyov, X. Zhu, J. Zhu and C. Chien, *Phys Rev Lett* **96,** 027205 (2006)

[3] J. Zhu, Y. Zheng and G. Prinz, *J. Appl. Phys.* **87,** 6668 (2000)

[4] M. Klaui, J. Rothman, L. Lopez-Diaz, C. Vaz, J. Bland and Z. Cui, *Appl Phys Lett* **78,** 3268 (2001)

[5] W. Meiklejohn and C. Bean, *Phys Rev* **102,** 1413 (1956)

[6] M. Stiles and R. McMichael, *Phys Rev B* **59,** 3722 (1999)

[7] Z. Li, O. Petracic, R. Morales, J. Olamit, X. Batlle, K. Liu and I. Schuller, *Phys Rev Lett* **96,** 217205 (2006)

[8] D. Suess, M. Kirschner, T. Schrefl, J. Fidler, R. Stamps and J. Kim, *Phys Rev B* **67,** 054419 (2003)

[9] T. Ambrose, R. Sommer and C. Chien, *Phys Rev B* **56,** 83 (1997)

[10] M. Gruyters and D. Riegel, *Phys Rev B* **63** 052401 (2000)

[11] M. Kiwi, J. Mejia-Lopez, R. Portugal and R. Ramirez, *Solid State Commun.* **116,** 315 (2000)

[12] J. Sort, K. Buchanan, V. Novosad, A. Hoffmann, G. Alvarez, A. Bollero, M. Baro, B. Dieny and J. Nogues, *Phys Rev Lett* **97,** 067201 (2006)

[13] D. Singh, R. Krotkov, H. Xiang, T. Xu, T. Russell and M. Tuominen, *Nanotechnology* **19,** 245305 (2008)

[14] W. Meiklejohn and C. Bean, *Phys Rev* **105,** 904 (1957)

[15] A. Berkowitz and K. Takano, *J Mag Mag Mat* **200,** 552 (1999)

[16] R. McMichael, C. Lee, M. Stiles, F. Serpa, P. Chen and W. Egelhoff, *Journal of Appl Phys* **87,** 6406 (2000)

[17] J. Nogues, D. Lederman, T. Moran and I. Schuller, *Phys Rev Lett* **76,** 4624 (1996)

[18] A. Malozemoff, *Phys Rev B* **35,** 3679 (1987)

[19] J. Coey, Chapter **12** "Spin Electronics" (2000), *Springer Publishing Group*

[20] K. Ounadjela, I. Prejbeanu, L. Buda, U. Ebel and M. Hehn, Chapter **15**, "Spin-Electronics" (2000), *Springer Publishing Group*

[21] A. Ercole, E. Kernohan, G. Lauho and J. Bland, *J. Mag. Mag. Mat.* **198-199,** 534 (1999)

[22] M. Kiwi, J. Lopez, R. Portugal and R. Ramirez, *Europhys. Lett.* **48**(5), 573 (1999)

[23] M. Bertotti, "Hysteresis and Magnetism", Academic Press (1998)